\documentclass[pra,twocolumn,aps,floatfix,showpacs,tightenlines,superscriptaddress]{revtex4-1}

\usepackage{amssymb,amsmath,amsthm,amsbsy,epsfig,color,graphicx,times}
\usepackage{epstopdf}
\usepackage{stackengine}

\def\ket#1{|{#1}\rangle}

\begin{document}

\title{Classical nature of ordered quantum phases and origin of spontaneous symmetry breaking}

%\title{Entanglement and quantum correlations in many-body systems: \\
%       a unified approach via local unitary operations}

\author{M. Cianciaruso}
\affiliation{School of Mathematical Sciences, The University of Nottingham, University Park, Nottingham NG7 2RD, United Kingdom}
\affiliation{Dipartimento di Fisica ``E. R. Caianiello'', Universit\`a degli Studi di Salerno,
Via Giovanni Paolo II 132, I-84084 Fisciano (SA), Italy}
\affiliation{INFN Sezione di Napoli, Gruppo collegato di Salerno, Italy}

\author{S. M. Giampaolo}
%\affiliation{University of Vienna, Faculty of Physics, Boltzmanngasse 5, 1090 Vienna, Austria}
\affiliation{International Institute of Physics, UFRN, Av. Odilon Gomes de Lima 1722, 59078-400 Natal, Brazil}
\affiliation{Dipartimento di Ingegneria Industriale, Universit\`a degli Studi di Salerno, Via Giovanni Paolo II 132, I-84084 Fisciano (SA), Italy}

\author{L. Ferro}
\affiliation{Dipartimento di Ingegneria Industriale, Universit\`a degli Studi di Salerno, Via Giovanni Paolo II 132, I-84084 Fisciano (SA), Italy}
\affiliation{INFN Sezione di Napoli, Gruppo collegato di Salerno, Italy}

\author{W. Roga}
\affiliation{Department of Physics, University of Strathclyde, John Anderson Building, 107 Rottenrow, Glasgow G4 0NG, United Kingdom}
\affiliation{Dipartimento di Ingegneria Industriale, Universit\`a degli Studi di Salerno, Via Giovanni Paolo II 132, I-84084 Fisciano (SA), Italy}

\author{G. Zonzo}
\affiliation{Dipartimento di Fisica ``E. R. Caianiello'', Universit\`a degli Studi di Salerno, Via Giovanni Paolo II 132, I-84084 Fisciano (SA), Italy}
\affiliation{Dipartimento di Ingegneria Industriale, Universit\`a degli Studi di Salerno, Via Giovanni Paolo II 132, I-84084 Fisciano (SA), Italy}

\author{M. Blasone}
\affiliation{Dipartimento di Fisica ``E. R. Caianiello'', Universit\`a degli Studi di Salerno,
Via Giovanni Paolo II 132, I-84084 Fisciano (SA), Italy}
\affiliation{INFN Sezione di Napoli, Gruppo collegato di Salerno, Italy}

\author{F. Illuminati}
\thanks{Corresponding author: filluminati@unisa.it}
\affiliation{Dipartimento di Ingegneria Industriale, Universit\`a degli Studi di Salerno, Via Giovanni Paolo II 132, I-84084 Fisciano (SA), Italy}
\affiliation{INFN Sezione di Napoli, Gruppo collegato di Salerno, Italy}
\affiliation{Consiglio Nazionale delle Ricerche, Istituto di Nanotecnologia, Rome Unit, I-00195 Roma, Italy}
%%\affiliation{CNISM Unit\`a di Salerno, I-84084 Fisciano (SA), Italy}

\date{April 21, 2016}

\begin{abstract}
We analyse the nature of spontaneous symmetry breaking in complex quantum systems by investigating the long-standing conjecture that the maximally symmetry-breaking quantum ground states are the most classical ones corresponding to a globally ordered phase. We make this argument quantitatively precise by comparing different local and global indicators of classicality and quantumness, respectively in symmetry-breaking and symmetry-preserving quantum ground states. We first discuss how naively comparing local, pairwise entanglement and discord apparently leads to the opposite conclusion. Indeed, we show that in symmetry-preserving ground states the two-body entanglement captures only a modest portion of the total two-body quantum correlations, while, on the contrary, in maximally symmetry-breaking ground states it contributes the largest amount to the total two-body quantum correlations. We then put to test the conjecture by looking at the global, macroscopic correlation properties of quantum ground states. We prove that the ground states which realize the maximum breaking of the Hamiltonian symmetries, associated to a globally ordered phase, are the only ones that: I) are always locally convertible, i.e. can be obtained from all other ground states by only applying LOCC transformations (local operations and classical communication), while the reverse is never possible; II) minimize the monogamy inequality on the globally shared, macroscopic bipartite entanglement.
\end{abstract}

%\begin{abstract}
%Local unitary operations allow for a unifying approach to the quantification of quantum correlations among the constituents of a bipartite quantum system. For pure states, the distance between a given state and its image under least-perturbing local unitary operations is a {\em bona fide} measure of quantum entanglement, the so-called entanglement of response, which can be extended to mixed states via the convex roof construction. On the other hand, when defined directly on mixed states perturbed by local unitary operations, such a distance turns out to be a {\em bona fide} measure of quantum correlations, the so-called discord of response. Exploiting this unified framework, we perform a detailed comparison between two-body entanglement and two-body quantum discord in infinite $XY$ quantum spin chains both in symmetry-preserving and symmetry-breaking ground states as well as in thermal states at finite temperature. The results of the investigation show that in symmetry-preserving ground states the two-point quantum discord dominates over the two-point entanglement, while in symmetry-breaking ground states the two-point quantum discord is strongly suppressed and the two-point entanglement is essentially unchanged. In thermal states, for certain regimes of Hamiltonian parameters, we show that the pairwise quantum discord and the pairwise entanglement can increase with increasing thermal fluctuations.
%\end{abstract}

\pacs{03.67.Mn, 03.65.Ud, 75.10.Pq, 05.30.Rt}

\maketitle

\section{Introduction}

In the study of collective quantum phenomena, the understanding of the globally ordered phases associated to local order parameters relies on the key concept of spontaneous symmetry breaking~\cite{Goldstone1962}. The latter is required to explain the existence of locally inequivalent ground states
that are not eigenstates of one or more symmetry operators for the corresponding many-body Hamiltonian~\cite{Sachdev2000}. In recent years, knowledge of quantum phase transitions has been sharpened by the application of methods and techniques originally developed in the field of quantum information~\cite{Amico2008,Verstraete2008}. Various types of quantum phase transitions have been indeed characterized by identifying the singular points in the derivatives of different measures of bipartite~\cite{Osterloh2002,Osborne2002} and multipartite entanglement~\cite{GiampaoloHiesmayr2013,Giampaolo2014}. Moreover, different ordered phases have been identified by looking at the factorization properties of different ground states~\cite{Giampaolo2008,Giampaolo2009,Giampaolo2010} or by studying the behavior of the ground-state fidelity under local or global variations of the Hamiltonian parameters~\cite{Zanardi2006,Zanardi2007}.

Efforts have been devoted to the investigation of the behavior of the bipartite concurrence~\cite{Osterloh2006}, multipartite
entanglement~\cite{deOliveira2006,GiampaoloHiesmayr2013,Giampaolo2014}, and quantum discord~\cite{Amico2012,Tomasello2012} for some specific symmetry-breaking ground states. However, on the whole, the complete understanding of the physical mechanism that selects the symmetry-breaking ground states in the thermodynamic limit remains an open problem~\cite{Bratteli2012,Arodz2012}.
In complete analogy with the case of classical phase transitions driven by temperature, the common explanation of this phenomenon invokes the unavoidable presence of some local, however small, perturbing external field that selects one of the maximally symmetry-breaking ground states (MSBGSs) among all the elements of the quantum ground space. Crucially, in this type of reasoning it is assumed that the MSBGSs are the most classical ones and thus the ones that are selected in real-world situations, under the effect of decoherence that quickly destroys macroscopic coherent superpositions.

At first glance, this notion appears to be obvious. For instance, in the paradigmatic case of the quantum Ising model, the ground space of the ferromagnetic phase at zero transverse field $h$ is spanned by two orthogonal product states $|0\rangle^{\otimes N}$ and $|1\rangle^{\otimes N}$ which are in the same class of pointer states of the typical decoherence argument, while the symmetric states $\Psi_{\pm} = 1/\sqrt{2}(|0\rangle^{\otimes N} \pm |1\rangle^{\otimes N})$ realize macroscopic coherent superpositions (Schroedinger cats) that are not stable under decoherence~\cite{Zurek2003,vanWezel2008}. Therefore, at zero transverse field $h$,
the situation is very clear: the only stable states are those that maximally break the symmetry of the Hamiltonian, and at the same time,
those that feature vanishing macroscopic total correlations, including entanglement, between spatially separated regions.

On the other hand, as we turn on the external field $h$, we have a whole range of values where, before a critical value $h=h_c$ is reached, there is a
magnetic order associated to spontaneous symmetry breaking~\cite{BM1971}, and the decoherence argument applies within the entire, globally ordered
phase. This means that, again, the only stable states are those that maximally break the Hamiltonian symmetry.
%As the field $h$ is turned on, the symmetry-breaking states are the only metastable ones, in close analogy with pointer states.
However, now the symmetry-breaking states are entangled, and their mixed-state reductions on arbitrary subsystems possess in general
nonvanishing pairwise entanglement~\cite{Osborne2002,Osterloh2002,Amico2008}, as well as pairwise quantum~\cite{Tomasello2012,Campbell2013}
and classical correlations~\cite{BM1971}. It is thus now unclear if and in what sense the MSBGSs are the most classical among all quantum
ground states. Indeed, as we shall see in the following, the symmetry-breaking ground states can be, in general, locally more entangled than
symmetric ground states (see also Ref.~\cite{Osterloh2006}). On the other hand, it is always implicitly assumed that such symmetry-breaking states are not
macroscopically correlated, while their symmetric superpositions are, in complete analogy with the case $h=0$. Although this is
a very plausible picture, a rigorous proof has never been provided, due to the mathematical difficulties in dealing with measures of macroscopic entanglement
and correlations based on the von Neumann entropy; see, e.g., the difficulties in proving the boundary ({\em area}) law in generic gapped
systems~\cite{arealaw1,arealaw2}. The symmetry-breaking states obey the boundary law for
entanglement~\cite{Holzhey1994,Vidal2003,Korepin2004}, while the macroscopic correlations featured by the superposition of
two different symmetry broken sectors are of order one. The task is then, to identify quantities that are able to distinguish the presence of macroscopic entanglement and quantum correlations, among all possible sources of entanglement and correlations.

In the present work we promote such qualitative picture to an explicit quantitative investigation on the nature of globally ordered quantum phases and the origin of spontaneous symmetry breaking, and we carry it out by comparing various quantifiers of local and global quantum correlations in symmetry-breaking and symmetry-preserving quantum ground states. We will first compare measures of local, pairwise quantum correlations and show that in symmetry-preserving ground states the two-body entanglement captures only a modest portion of the local, two-body quantum correlations, while in maximally symmetry-breaking ground states it accounts for the largest contribution. Next, we will introduce (see below) proper criteria and quantifiers of the degree of classicality of quantum states with respect to their global contents of macroscopic entanglement and quantum correlations. Finally, we will show that, within the quantum ground space corresponding to macroscopically ordered phases with nonvanishing local order parameters, the MSBGSs are the most classical ground states in the sense that they are the only quantum ground states that satisfy the following two criteria for each set of Hamiltonian parameters consistent with an ordered quantum phase in the thermodynamic limit:
\begin{itemize}
\item {\em Local convertibility} -- All global ground states are convertible into MSBGSs applying only local operations and classical communication (LOCC transformations), while the reverse transformation is impossible;
\item {\em Entanglement distribution} -- The MSBGSs are the only global ground states that minimize the residual tangle between a dynamical variable and the remainder of a macroscopic quantum system. Stated otherwise, the MSBGSs are the only ground states that satisfy the monogamy inequality -- a strong constraint, with no classical counterpart, on the shared bipartite entanglement between all components of a macroscopic quantum system -- at its minimum among all other possible ground states, and thus minimize the macroscopic multipartite entanglement as measured by the residual tangle.
\end{itemize}
Verification of these two features amounts to proving that the mechanism of spontaneous symmetry breaking selects the most classical ground states associated to globally ordered phases of quantum matter with nonvanishing local order parameters.

Our results are of general validity for all systems that belong to the same universality class of exactly solvable models that are standard prototypes for quantum phase transitions associated to spontaneous symmetry breaking, such as the $XY$ quantum spin models~\cite{Sachdev2000}.

The paper is organized as follows. In Section~\ref{sec:XYmodel} we recall the main features of the one-dimensional $XY$ models in transverse field with periodic boundary conditions. In Section~\ref{sec:definitionsandnotationsofstellarentandquant} we perform the comparison
between entanglement and discord for spin pairs in infinite $XY$ chains (thermodynamic limit), respectively in symmetry-preserving and MSBGSs. In Section~\ref{sec:symmetrybreakingorigin} we compare global (as opposed to pairwise) measures of classicality and quantumness, such as local convertibility and entanglement distribution, for symmetry-breaking and symmetry-preserving quantum ground states.  Conclusions and outlook are discussed in Section~\ref{sec:conclusions}.

\section{$XY$ MODELS}\label{sec:XYmodel}

The one-dimensional spin-$1/2$ $XY$ Hamiltonian with ferromagnetic nearest-neighbor interactions in a transverse field with periodic boundary conditions reads~\cite{Lieb1961,Pfeuty1970,Barouch1970,BM1971,Johnson1971}:
\begin{equation}\label{eq:XYmodelhamiltonian}
H \! =\!-\!\sum_{i=1}^{N}\! \! \left[\!\left(\!\frac{1+\gamma }{2}\!\right) \! \sigma_i^x \sigma_{i+1}^x \!+\!
\left(\!\frac{1-\gamma }{2}\!\right)\!\sigma_i^y
\sigma_{i+1}^y \!+\! h \sigma_i^z\right]\! \;,
\end{equation}
where $\sigma_i^\mu$, $\mu = x, y, z$, are the Pauli spin-$1/2$ operators acting on site $i$, $\gamma$ is the
anisotropy parameter in the $xy$ plane, $h$ is the transverse magnetic field, and the periodic boundary conditions
$\sigma_{N+1}^\mu \!\equiv\! \sigma_1^\mu$ ensure the invariance under spatial translations.

For this class of models, the phase diagram can be determined exactly in great detail~\cite{Lieb1961,Barouch1970}. In the thermodynamic limit, for any $\gamma\!\in\!(0,1]$, a quantum phase transition occurs at the critical value $h_c = 1$ of the transverse field. For $h\! <\! h_c\!=\!1$ the system is ferromagnetically ordered and is characterized by a twofold ground-state degeneracy such that the $\mathbb{Z}_2$ parity symmetry under inversions along the spin-$z$ direction is broken by some elements of the ground space. Given the two symmetric ground states,  the so-called even $|e\rangle$ and odd $|o\rangle$ states belonging to the two orthogonal subspaces associated to the two possible distinct eigenvalues of the parity operator, any symmetry-breaking linear superposition of the form
\begin{equation}\label{eq:groundstates}
|g(u,v)\rangle = u |e\rangle + v |o\rangle \;
\end{equation}
is also an admissible ground state, with the complex superposition amplitudes $u$ and $v$ constrained by the normalization condition \mbox{$|u|^2\!+\!|v|^2\!=\!1$}. Taking into account that the even and odd ground states are orthogonal, the expectation values of operators that commute with the parity operator are independent of the superposition amplitudes $u$ and $v$. On the other hand, spin operators that do not commute with the parity
may have nonvanishing expectation values on such linear combinations and hence break the symmetry of the Hamiltonian (\ref{eq:XYmodelhamiltonian}).
%This result and the following ones hold in general for all quantum models in the same universality class of the $XY$ Hamiltonian.

Consider observables $O_S$ that are arbitrary products of spin operators and anti-commute with the parity. Their expectation values in the superposition ground states
(\ref{eq:groundstates}) are of the form
\begin{equation}\label{eq:expectationvalue}
\langle g(u,v)|O_S|g(u,v)\rangle = u v^* \langle o|O_S|e\rangle + v u^* \langle e|O_S|o\rangle \; .
\end{equation}
Both $\langle o|O_S|e\rangle$ and $\langle e|O_S|o\rangle$ are real and independent of $u$ and $v$ and hence the expectation
(\ref{eq:expectationvalue}) is maximum for \mbox{$u\!=\!\pm v\!=\!1\!/\!\sqrt{2}$}~\cite{Barouch1970}. These are the values of the
superposition amplitudes that realize the maximum breaking of the symmetry and identify the order parameter as well as the MSBGSs.

Besides the quantum critical point, there exists another relevant value of the external magnetic field, that is $h_f=\sqrt{1-\gamma^2}$,
the {\em factorizing field}. Indeed, at this value of $h$, the system admits a two-fold degenerate, completely factorized ground
state~\cite{Kurmann1982,Roscilde2005,Giampaolo2008,Giampaolo2009,Giampaolo2010}.

In order to discuss the entanglement and discord-type correlations of quantum ground states, we consider arbitrary bipartitions $(A|B)$ such that subsystem $A=\{i_1,\ldots,i_L\}$ is any subset made of $L$ spins, and subsystem $B$ is the remainder. Given any global ground state of the total system, the reduced density matrix $\rho_A$ ($\rho_B$) of subsystem $A$ ($B$) can be expressed in general in terms of the $n$-point correlation functions~\cite{Osborne2002}:
\begin{equation}
\label{eq:defreduce}
 \rho_{A}\!(u,v) \!= \! \frac{1}{2^L} \!\!\!\!\!\!\!
 \ \sum_{\mu_1,\ldots,\mu_L}\! \!\!\!\! \langle g(u,v)\!| \sigma_{i_1}^{\mu_1}\! \cdots \! \sigma_{i_L}^{\mu_L} \!
 |g(u,v)\!\rangle \sigma_{i_1}^{\mu_1}\! \cdots \! \sigma_{i_L}^{\mu_L} \, ,
\end{equation}
and analogously for $\rho_B$. All expectations in Eq.~(\ref{eq:defreduce}) are associated to spin operators that either commute or anti-commute with the parity along the spin-$z$ direction. Therefore the reduced density matrix $\rho_A$ can be expressed as the sum of a symmetric part $\rho_A^{(s)}$, i.e. the reduced density matrix obtained from $|e\rangle$ or $|o\rangle$, and a traceless matrix $\rho_A^{(a)}$ that includes all the terms that are nonvanishing only in the presence of a breaking of the symmetry:
\begin{equation}
\label{eq:defreduce1}
 \rho_{A}(u,v) = \rho_A^{(s)} + (uv^*+vu^*) \rho_A^{(a)} \; .
\end{equation}
Both $\rho_A^{(s)}$ and $\rho_A^{(a)}$ are independent of the superposition amplitudes $u$ and $v$, while the reduced density matrix depends on the choice of the ground state. This implies that the elements of the ground space are not locally equivalent. Explicit evaluation of expectations and correlations in symmetry-breaking ground states in the thermodynamic limit is challenging even when the exact solution for the symmetric elements of the ground space is available.

We will now sketch a method that allows to overcome this difficulty and whose general validity is not in principle restricted to the particular model considered. In order to obtain $\rho_A^{(s)}$ it is sufficient to transform the spin operators in fermionic ones and then apply Wick's theorem. Such algorithm cannot be applied to spin operators $O_A$, acting on subsystem $A$, that anti-commute with the parity. In order to treat this case we first introduce the symmetric operator $O_AO_{A+r}$, for which, by applying the previous procedure, we can evaluate $\langle e |O_A O_{A+r}| e\rangle$. Then, the desired expectation $\langle e |O_A| o\rangle$ can be computed by exploiting the property of asymptotic factorization of products of local operators at infinite separation~\cite{Barouch1970,Sachdev2000,Bratteli2012} that yields $\langle e |O_A| o\rangle = \sqrt{\lim\limits_{r \to \infty} \langle e |O_A O_{A+r}| e\rangle}$,
where the root's sign is fixed by imposing positivity of the density matrix $\rho_{A}(u,v)$. Having obtained the exact reduced density matrix $\rho_{A}(u,v)$ for all possible subsystems $A$ and superposition amplitudes $u$ and $v$, we are equipped to investigate the nature of quantum ground states with respect to their properties of classicality and quantumness.

\section{Two-body quantum correlations}\label{sec:definitionsandnotationsofstellarentandquant}

In this Section we analyze the behavior of one-way discord-type correlations and entanglement between any two spins for different ground states. One-way discord-type correlations are properties of quantum states more general than entanglement. Operationally, they are defined in terms of state distinguishability with respect to the so-called {\em classical-quantum} states. The latter are quantum states that, besides being separable, i.e. not entangled, remain invariant under the action of at least one nontrivial local unitary operation. In geometric terms, a {\em bona fide} measure of quantum correlations must quantify how much a quantum state {\em discords} from classical-quantum states and must be invariant under the action of all local unitary operations. A computable and operationally well defined geometric measure of quantum correlations is then the {\em discord of response}~\cite{Roga2014,Giampaolo2013}. The pairwise discord of response $D_R$ for a two-spin reduced density matrix is defined as:
\begin{equation}
\label{discord}
D_R(\rho_{ij}^{(r)}(u,v)) \equiv \frac{1}{2} \min_{U_i} d_{x} \left(\rho_{ij}^{(r)}(u,v),\tilde{\rho}_{ij}^{(r)}(u,v) \right)^2 \, ,
\end{equation}
where $\rho_{ij}^{(r)}(u,v)$ is the state of two spins $i$ and $j$ at a distance $r$, obtained by taking the partial trace of
the ground state $|g(u,v)\rangle$ with respect to all other spins in the system, $\tilde{\rho}_{ij}^{(r)}(u,v)\!\equiv\! U_i\rho_{ij}^{(r)}(u,v) U_i^\dagger$ is the two-spin state transformed under the action of a local unitary operation $U_i$ acting on spin $i$, and $d_{x}$ is any well-behaved, contractive distance (e.g. Bures, trace, Hellinger) of $\rho_{ij}^{(r)}$ from the set of locally unitarily perturbed states, realized by the least-perturbing operation in the set. The trivial case of the identity is excluded by considering only unitary operations with {\em harmonic} spectrum, i.e. the fully non-degenerate spectrum on the unit circle with equispaced eigenvalues.

For pure states the discord of response reduces to an entanglement monotone, whose convex-roof extension to mixed states is the so-called \textit{entanglement of response}~\cite{Giampaolo2007,Monras2011,Gharibian2012}. Therefore, the entanglement and the discord of response quantify different aspects of bipartite quantum correlations via two different uses of local unitary operations. The discord of response arises by applying local unitaries directly to the generally mixed state, while the entanglement of response stems from the application of local unitaries to pure states. By virtue of their common origin, it is thus possible to perform a direct comparison between these two quantities.

In terms of the trace distance, which will be relevant in the following, the two-qubit entanglement of response is simply given by the squared concurrence~\cite{Wootters1998,Roga2014}, whereas the two-qubit discord of response relates nicely to the trace distance-based geometric discord~\cite{Nakano2013}, whose closed formula is known only for a particular class of two-qubit states~\cite{Ciccarello2014}, although it can be computed for a more general class of two-qubit states through a very efficient numerical optimization.

\subsection{Symmetry-preserving ground states}\label{sec:symmetricgroundstate}

\begin{figure}[t]
\includegraphics[width=8cm]{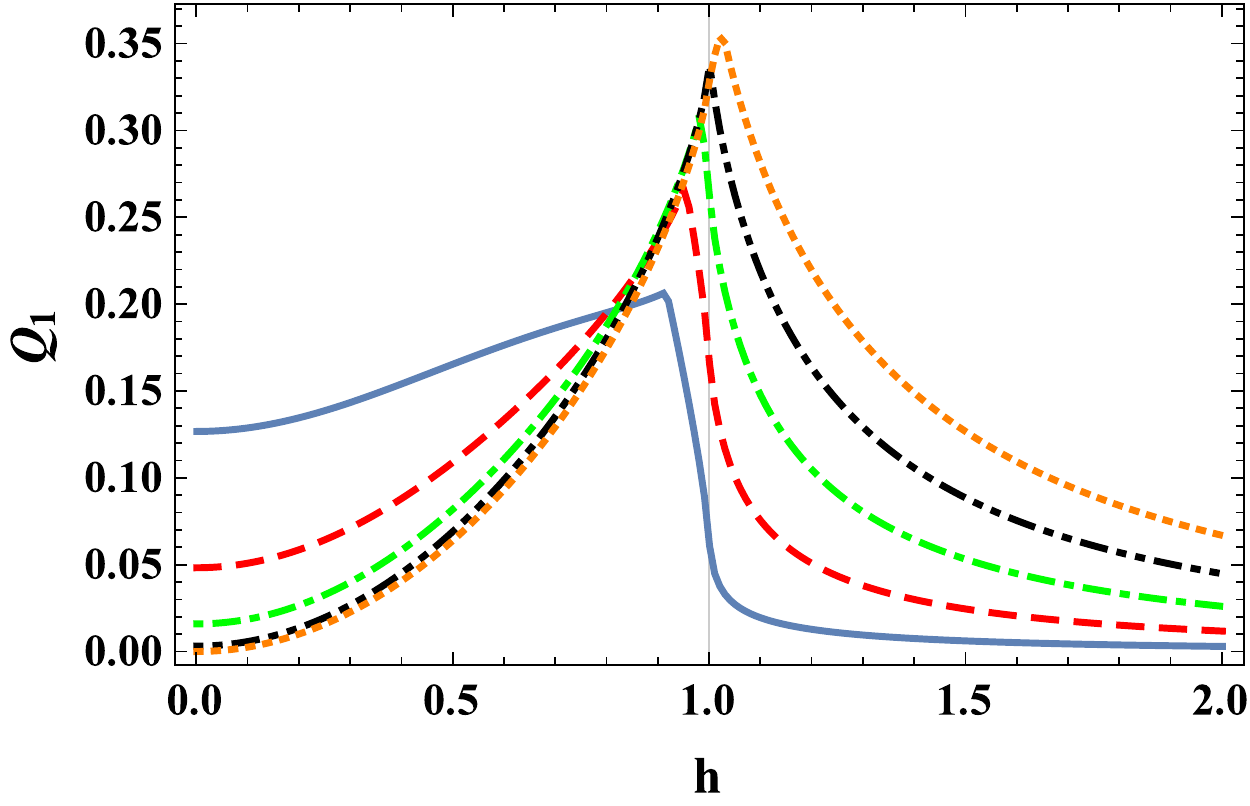}
\includegraphics[width=8cm]{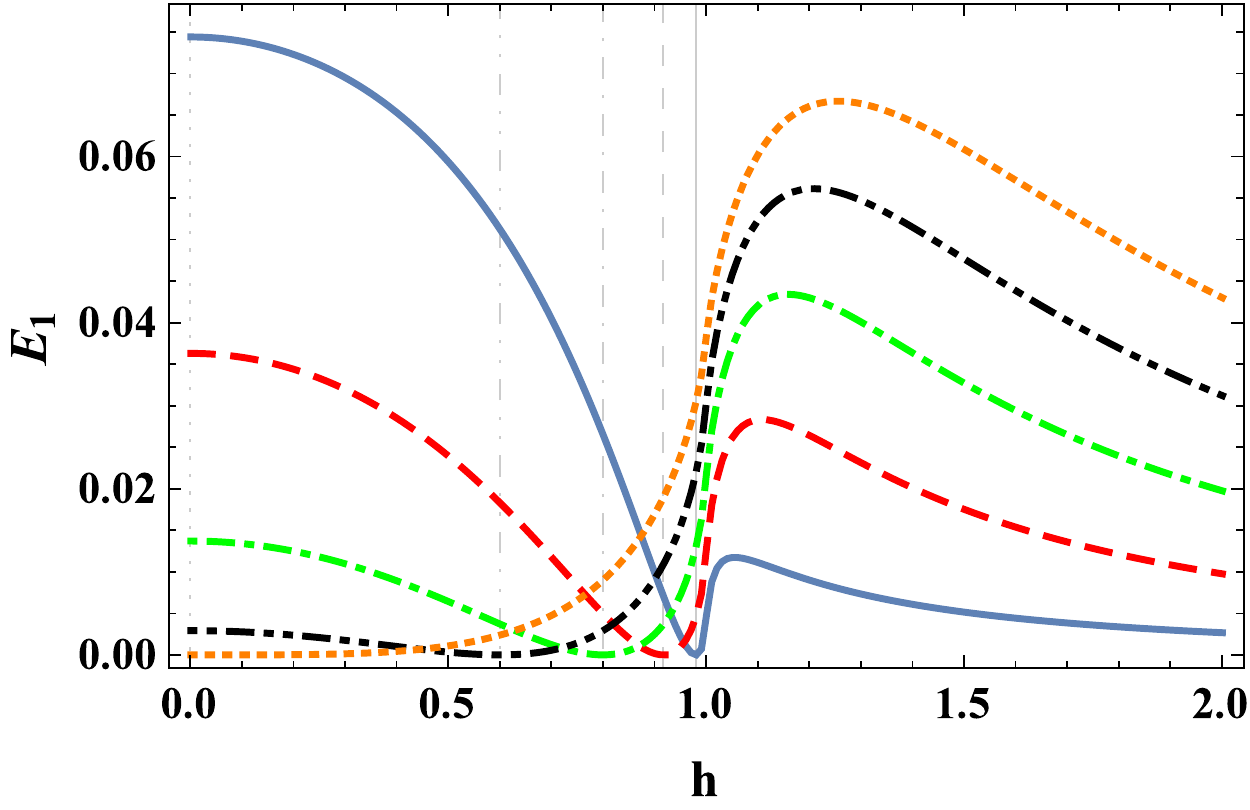}
\caption{Nearest-neighbor trace distance-based discord of response (upper panel) and nearest-neighbor trace distance-based entanglement of response (lower panel) for symmetry-preserving ground states, in the thermodynamic limit, as functions of the external field $h$, and for different values of the anisotropy $\gamma$. Solid blue curve: $\gamma=0.2$; dashed red curve: $\gamma=0.4$; dot-dashed green curve: $\gamma=0.6$; double-dot-dashed black curve: $\gamma=0.8$; dotted orange curve: $\gamma=1$. In the lower panel, to each of these curves, there corresponds a vertical line denoting the associated factorizing field $h_f$. In the upper panel, the solid vertical line denotes the critical field $h_c = 1$.}
\label{fig:symnearestquantandent}
\end{figure}

We first compare the two-body entanglement of response and the two-body discord of response in symmetry-preserving ground states.
For two neighboring spins, these two quantities are plotted in Fig.~\ref{fig:symnearestquantandent} as functions of the external field $h$ and for different values of the anisotropy $\gamma$. For any intermediate value of $\gamma$, the nearest-neighbor entanglement of response $E_1$ exhibits the following behavior. If $h<h_f$, $E_1$ decreases until it vanishes at the factorizing field $h=h_f$. Otherwise, if $h>h_f$, $E_1$ first increases until it reaches a maximum at some value of $h$ higher than the critical point $h_c=1$, then it decreases again until it vanishes asymptotically for very large values of $h$ in the paramagnetic phase (saturation). Overall, $E_1$ features two maxima at $h=0$ and $h>h_c$ and two minima at $h=h_f$ (factorization) and $h\rightarrow\infty$ (saturation). For the Ising model ($\gamma=1$) the point $h=0$ corresponds instead to a minimum,
since it coincides with the factorizing field $h_f=\sqrt{1-\gamma^2}$.
%while in the isotropic $XX$ model ($\gamma=0$) there is no second maximum for large fields $h>h_c$, since the ground state is always completely factorized as soon as $h \ge h_c$.

On the other hand, regardless of the value of $\gamma$, the nearest-neighbor discord of response $Q_1$ always features a single maximum. Depending on the value of $\gamma$ such maximum can be either in the ordered phase $h<h_c$ or in the disordered (paramagnetic) phase $h>h_c$, moving towards higher values of $h$ with increasing $\gamma$. Remarkably, $Q_1$ never vanishes at the factorizing field, except in the extreme case of $\gamma=1$. Indeed, at the factorizing field $h=h_f$ and for any $\gamma\neq 0,1$, the symmetry-preserving ground state is not completely factorized but rather
is a coherent superposition with equal amplitudes of the two completely factorized MSBGSs. Consequently, while the two-body entanglement of response must vanish in accordance with the convex roof extension, the two-body discord of response remains always finite.

\begin{figure}[t]
\includegraphics[width=7.65cm]{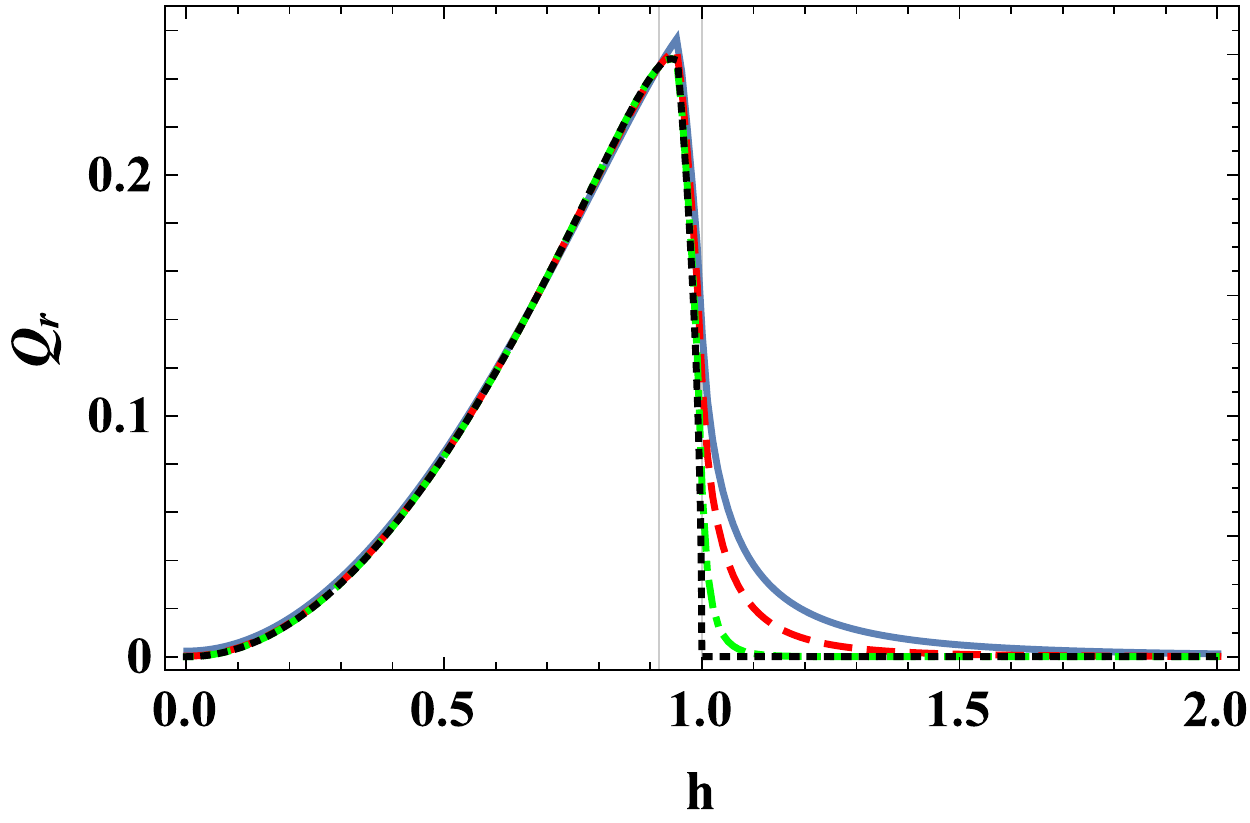}
\includegraphics[width=7.65cm]{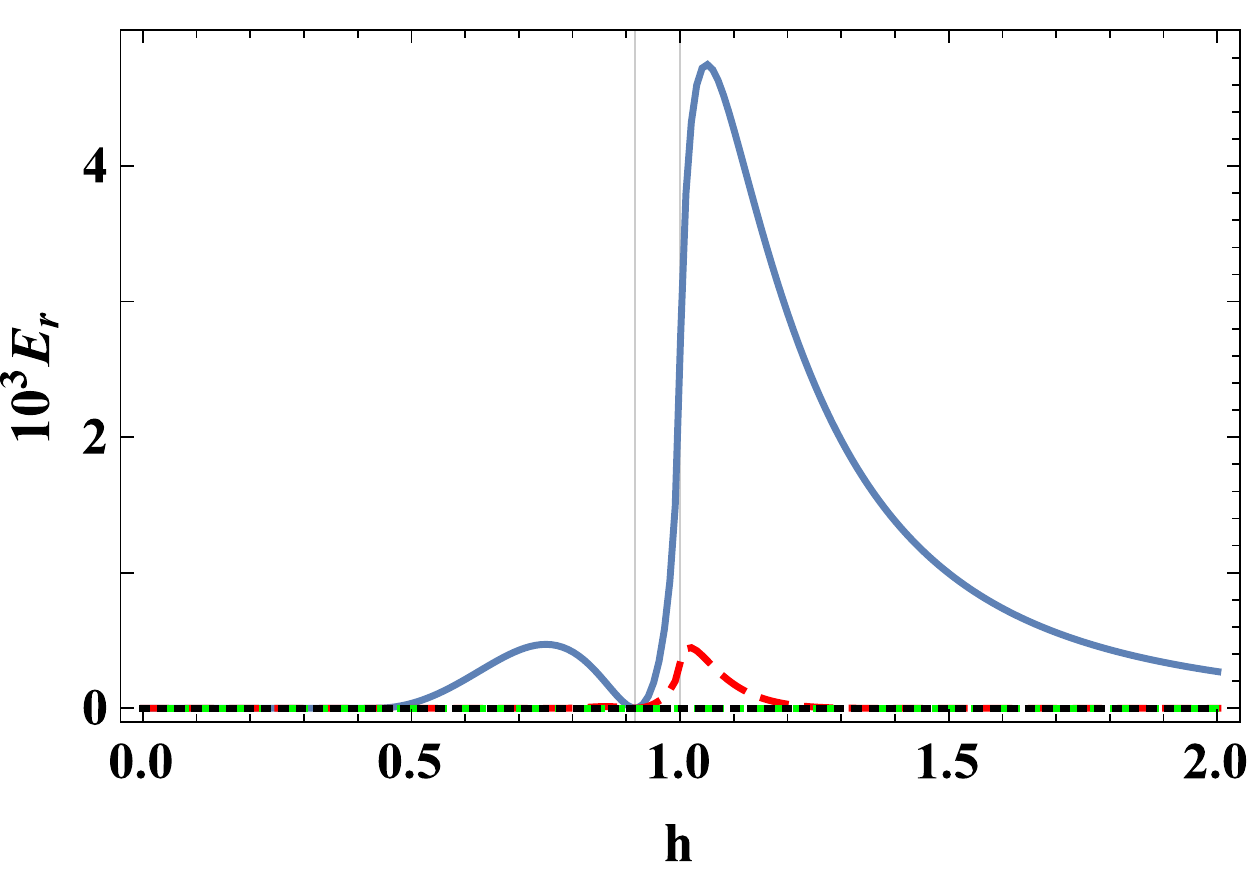}
\caption{Two-body trace distance-based discord of response (upper panel) and two-body trace distance-based entanglement of response (lower panel) for symmetry-preserving ground states, in the thermodynamic limit, as functions of the external field $h$, in the case of $\gamma=0.4$, for different inter-spin distances $r$. Solid blue curve: $r=2$; dashed red curve: $r=3$; dot-dashed green curve: $r=8$; dotted black curve: $r=\infty$. In both panels, the two solid vertical lines correspond, respectively, to the factorizing field (left) and to the critical field (right).}
%Inset (both panels): same, but with $\gamma=0$; the solid vertical line corresponds to the critical point.}
%two-body (a) discord of response and (b) entanglement of response for symmetry-preserving ground states, in the thermodynamic limit, as functions of the external field $h$, at $\gamma=0$, and for different distances $r$ between the two spins: solid/blue line $r=2$; dashed/red line $r=3$; dot-dashed/green line $r=8$; and dotted/black line $r=\infty$.
\label{fig:symnquantandentversushatvariousr}
\end{figure}

When increasing the inter-spin distance $r$, the pairwise entanglement of response $E_r$ and discord of response $Q_r$ behave even more differently (see Fig.~\ref{fig:symnquantandentversushatvariousr}). Due to the monogamy of the squared concurrence~\cite{Coffman2000,Osborne2006}, $E_r$ dramatically drops to zero as $r$ increases, except in a small region
around the factorizing field $h=h_f$ that gets smaller and smaller as $r$ increases, in agreement with the findings of Ref.~\cite{Amico2006}. On the other hand, while in the disordered and critical phases $Q_r$ vanishes as $r$ increases, in the ordered phase $Q_r$ survives even in the limit of
infinite $r$. Indeed, in both the disordered and critical phases, and when $r$ goes to infinity, the only non-vanishing one-body and two-body correlation functions in the symmetry-preserving ground states are
$\langle \sigma_i^z \rangle$ and $\langle \sigma_i^z \sigma_{i+r}^z\rangle$, so that the two-body reduced state can be written as
a classical mixture of eigenvectors of $\sigma_i^z \sigma_{i+r}^z$. On the other hand, in the ordered phase, also the two-body correlation function $\langle \sigma_i^x \sigma_{i+r}^x\rangle$ appears, while $\langle \sigma_i^x \rangle$ vanishes due to symmetry preservation, thus preventing the two-body marginal of the symmetry-preserving ground state from being a mixture of classical states.

\subsection{Maximally symmetry-breaking ground states}\label{sec:symmmetrybrokengroundstate}

\begin{figure}[t]
\includegraphics[width=8cm]{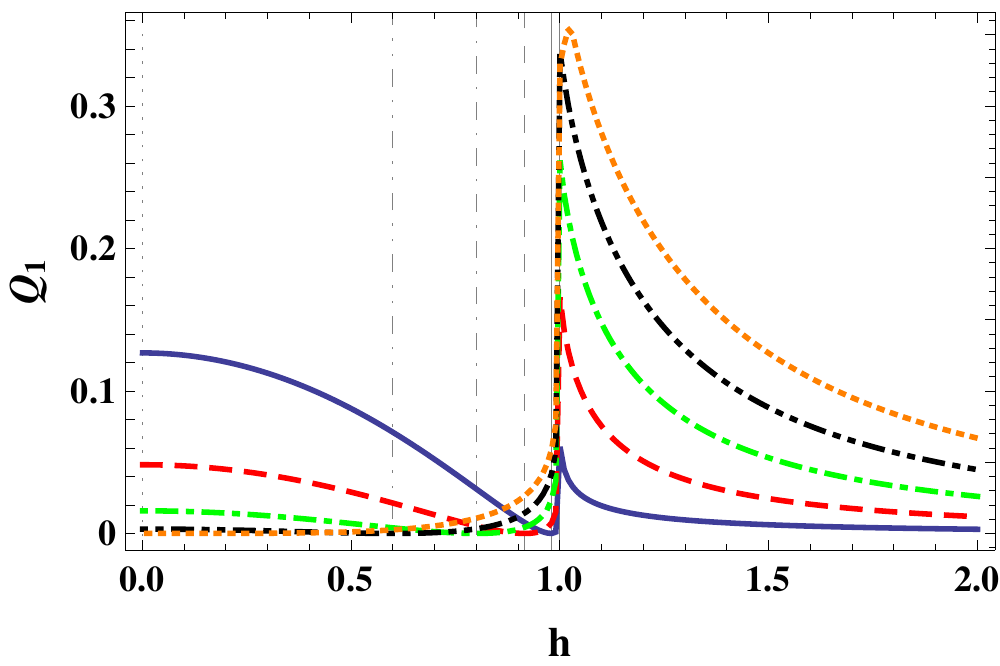}
\includegraphics[width=8cm]{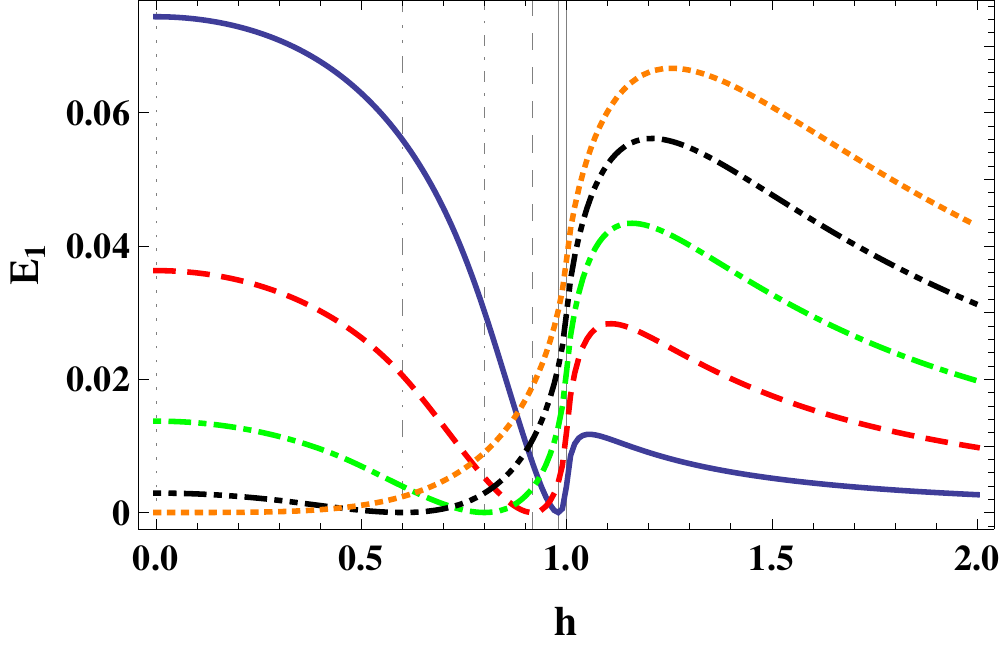}
\caption{Nearest-neighbor trace distance-based discord of response (upper panel) and nearest-neighbor trace distance-based entanglement of response (lower panel) in MSBGSs as  functions of the external field $h$, for different values of the anisotropy $\gamma$. Solid blue curve: $\gamma=0.2$; dashed red curve: $\gamma=0.4$;
dot-dashed green curve: $\gamma=0.6$; double-dot-dashed black curve: $\gamma=0.8$; dotted orange curve: $\gamma=1$. In both panels, to each of these curves, there corresponds a vertical line denoting the associated factorizing field $h_f$. The rightmost vertical line denotes the critical point.}
\label{fig:symbrokennearestquantandent}
\end{figure}
\begin{figure}[hbtp]
\includegraphics[width=8cm]{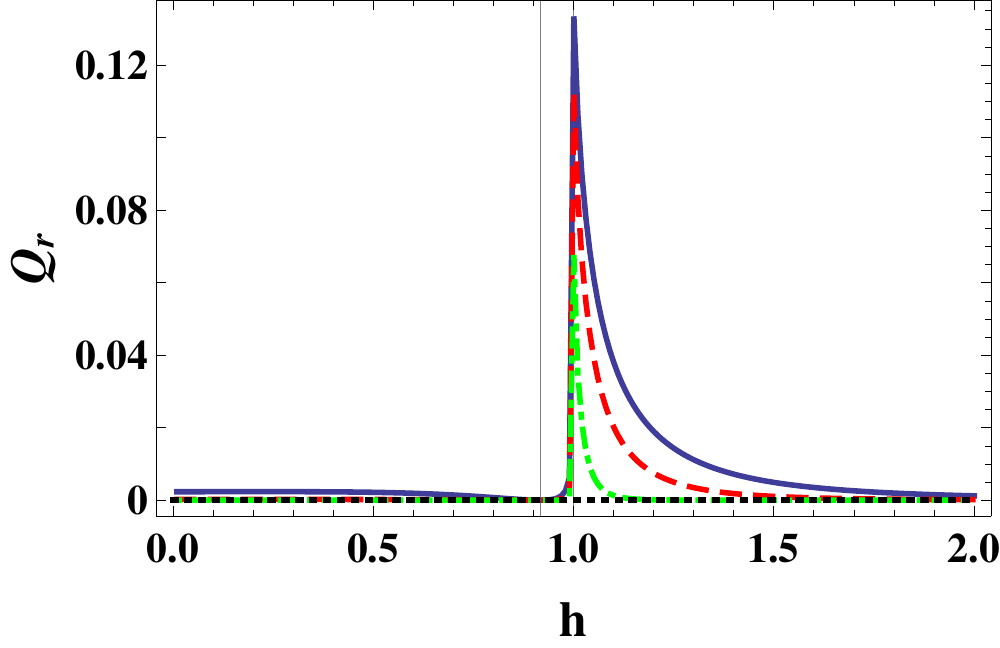}
\includegraphics[width=8cm]{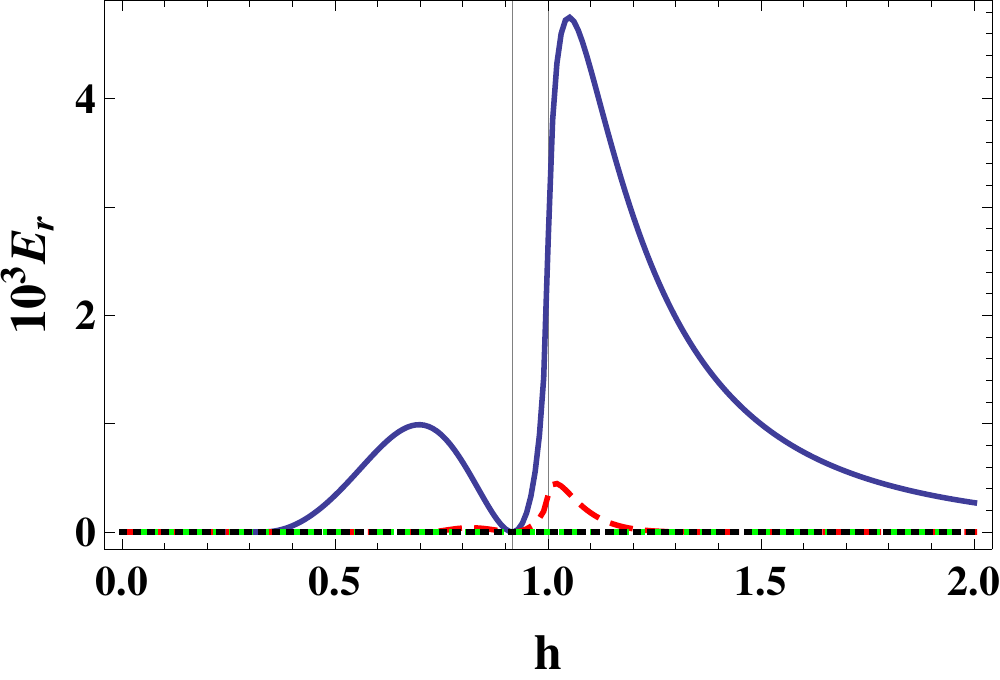}
\caption{Two-body trace distance-based discord of response (upper panel) and two-body trace distance-based entanglement of response (lower panel) in MSBGSs as functions of the external field $h$, at $\gamma=0.4$, for different inter-spin distances $r$. Solid blue curve: $r=2$; dashed red curve: $r=3$;
dot-dashed green curve: $r=8$; dotted black curve: $r=\infty$. In both panels, the two solid vertical lines correspond, respectively, to the factorizing field (left) and to the critical field (right).}
\label{fig:symbrokennquantandentversushatvariousr}
\end{figure}

In this section we move the focus of the comparison between two-body entanglement of response and discord of response from symmetry-preserving to MSBGSs. Spontaneous symmetry breaking manifests itself in the thermodynamic limit, in the ordered phase $h<h_c=1$ and for any non zero anisotropy $\gamma$, so that hereafter we will restrict the region of the phase space under investigation accordingly.

%The symmetry-preserving ground states depart from the symmetry-preserving ones due to the appearance of the non vanishing one-body correlation function $\left<\sigma^x_i\right>$ and the two-body correlation functions $\left<\sigma^x_i\sigma^z_j\right>$.

Fig.~\ref{fig:symbrokennearestquantandent} shows that, as soon as symmetry breaking is taken into account,
%the nearest-neighbor discord of response $Q_1$ becomes discontinuous at the critical point $h_c=1$, whereas the first derivative of the nearest-neighbor entanglement of response $\partial_h E_1$ still diverges logarithmically. In other words,
only the discord of response is affected by symmetry breaking at the critical point $h_c=1$. In fact, according to Ref.~\cite{Osterloh2006}, the concurrence and, consequently, the two-body entanglement of response, attain the same value for any $h\geq h_f$ both in the symmetry-preserving and MSBGSs. Otherwise, if $h<h_f$, there is a slight enhancement in the pairwise entanglement of response in the MSBGSs compared to the corresponding symmetry-preserving ones. Conversely, in general, the pairwise discord of response undergoes a dramatic suppression in the entire ordered phase $h<h_c$ when moving from symmetry-preserving to MSBGSs.

Considering the dependence on the inter-spin distance, we observe that the pairwise discord of response loses its long-range nature when moving from symmetry-preserving to MSBGSs (see Fig.~\ref{fig:symbrokennquantandentversushatvariousr}). More precisely, both the pairwise entanglement of response and the pairwise discord of response vanish asymptotically with increasing inter-spin distance. In the case of the pairwise entanglement of response, this result is again due to the monogamy of the squared concurrence~\cite{Coffman2000,Osborne2006}. In the case of the pairwise discord of response, it is instead due to the fact that not only the correlation function $\langle \sigma_i^x \sigma_{i+r}^x\rangle$ but also $\langle \sigma_i^x\rangle$ and $\langle \sigma_i^x \sigma_{i+r}^z\rangle$ are nonvanishing in the limit of infinite inter-spin distance $r$. This feature allows to write any two-spin reduced density matrix obtained from the MSBGSs as a classical mixture of eigenvectors of $O_i O_{i+r}$,
where $O_i$ is an Hermitian operator defined on the $i$-th site as $O_i= \cos \beta \sigma_i^z + \sin \beta \sigma_i^x$ with $\tan \beta= \frac{\langle \sigma_i^x\rangle}{\langle \sigma_i^z\rangle}$.

Overall, the quantum correlations between any two spins decrease significantly in the entire ordered phase when symmetry breaking is taken into account, and are almost entirely made up by contributions due to entanglement. In particular, at the factorizing field $h_f$, both the entanglement of response and the discord of response vanish. Indeed, we recall that the factorizing field $h_f$ owes its name to the two MSBGSs that are completely separable (product) at such value of the external magnetic field.

\section{Global properties: local convertibility and many-body entanglement sharing}\label{sec:symmetrybreakingorigin}

We now investigate the nature of quantum ground states in the ordered phase with respect to the properties of local convertibility of the global ground states and the many-body entanglement distribution.

\subsection{Local convertibility of many-body quantum ground states}

We begin by studying the property of local convertibility of quantum ground states in an ordered phase. In general, given two pure bipartite quantum states, $\ket{\psi_1}$ and $\ket{\psi_2}$, we say that $\ket{\psi_1}$ is locally convertible into $\ket{\psi_2}$ if $\ket{\psi_1}$ can be transformed into $\ket{\psi_2}$ by using only local quantum operations and classical communication (LOCC), and the aid of an ancillary entangled system~\cite{Jonathan1999,JonathanPlenio1999}.
%Entanglement is a key feature of quantum states with no classical counterpart. In particular, entanglement enjoys two basic properties that have no analogue %in classical correlations: monotonicity under local operations and classical communication (LOCC), and monogamy constraint on its shareability among many %parties.
%Let us first consider entanglement monotonicity and its consequences on the concept of local convertibility. The latter states that two pure states are %locally convertible  Based on the fact that quantum entanglement cannot increase under LOCC, it has been proven that a bipartite pure state can be converted %into a different one only by means of local operations plus a classical channel if and only if the entanglement of the initial state bounds from above the %entanglement of the transformed one

This concept of local convertibility can be formalized in terms of the entire hierarchy of the R\'enyi entanglement entropies
$S_\alpha(\rho_A) = \frac{1}{1-\alpha}\log_2\left[Tr(\rho_A^\alpha)\right]$
of the reduced density operator of subsystem $A$, which provides a complete characterization of the entanglement spectrum and its scaling behavior in different quantum phases~\cite{GiampaoloMontangero2013}. In a many-body setting, the necessary and sufficient conditions for a bipartite global state $\ket{\psi_1}$ to be locally convertible to another global state $\ket{\psi_2}$ is that the inequality $S_\alpha(\psi_1) \geq S_\alpha(\psi_2)$ holds for all bipartitions and all $\alpha>0$~\cite{Turgut2007}. Local convertibility has been recently applied to the characterization of topological order and the computational power of different quantum phases~\cite{HammaCincio2013,Cui2012,Cui2013}.
\begin{figure}[t]
\centering
\includegraphics[width=9cm]{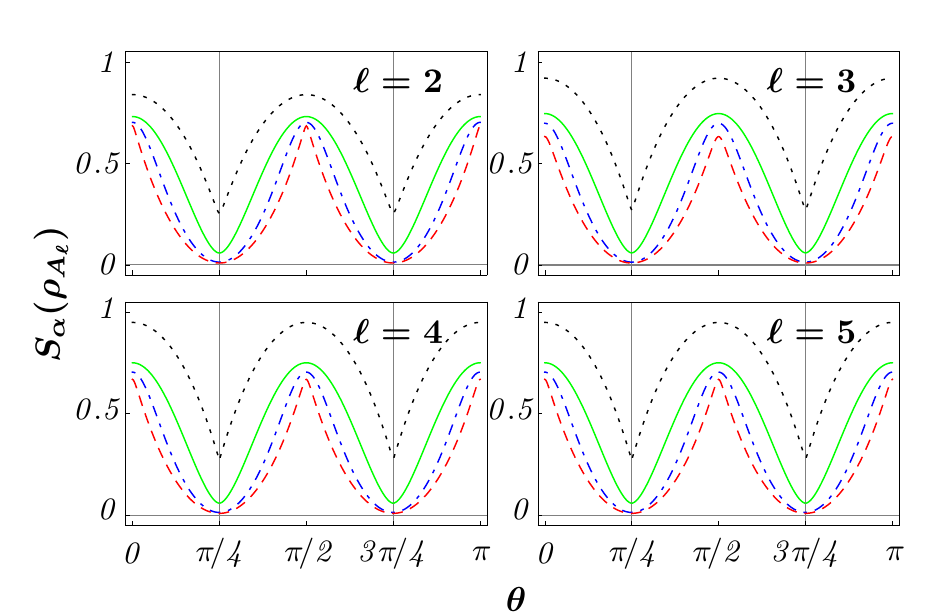}
\caption{Behavior of the R\'enyi entropies $S_\alpha(\rho_A)$ as functions of the different ground states in the ordered phase, $h < h_c$,
in the case of a subsystem $A_{\ell}$ made of $\ell$ contiguous spins. Each line stands for a different value of $\alpha$. Black dotted
line: $\alpha=0.5$. Green solid line: $\alpha\rightarrow 1^+$ (von Neumann entropy). Blue dot-dashed line: $\alpha=3$. Red dashed line:
$\alpha\rightarrow \infty$. The different ground states are parameterized by the superposition amplitudes $u=\cos(\theta)$ and $v=\sin(\theta)$. The two vertical lines correspond to the two MSBGSs, respectively obtained for $\theta=\pi/4$ and $\theta=3 \pi/4$. The Hamiltonian parameters are set at the intermediate values $\gamma=0.5$ and $h=0.5$. Analogous behaviors are observed for different values of the anisotropy and external field.}
\label{convertibilityversusell}
\end{figure}
\begin{figure}[t]
\centering
\includegraphics[width=9cm]{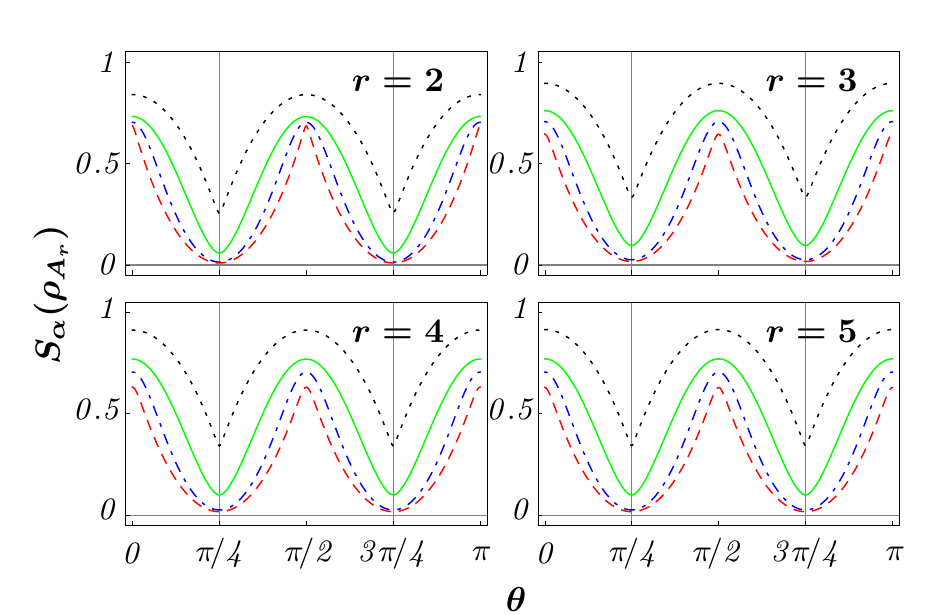}
\caption{Behavior of the R\'enyi entropies $S_\alpha(\rho_A)$ as functions of the different ground states in the ordered phase, $h < h_c$,
in the case of a subsystem $A_{r}$ made by two spins, for different inter-spin distances $r$. Each line stands for a different value of $\alpha$. Black dotted
line: $\alpha=0.5$. Green solid line: $\alpha\rightarrow 1^+$ (von Neumann entropy). Blue dot-dashed line: $\alpha=3$. Red dashed line:
$\alpha\rightarrow \infty$. The different ground states are parameterized by the superposition amplitudes $u=\cos(\theta)$ and $v=\sin(\theta)$. The two vertical lines correspond to the two MSBGSs, respectively obtained for $\theta=\pi/4$ and $\theta=3 \pi/4$. The Hamiltonian parameters are set at the intermediate values $\gamma=0.5$ and $h=0.5$. Analogous behaviors are observed for different values of the anisotropy and external field.}
\label{convertibilityversusr}
\end{figure}

It was previously shown that symmetric ground states are always locally convertible among themselves for $h_f < h < h_c$, and never for $h < h_f < h_c$~\cite{GiampaoloMontangero2013}. Here, thanks to the general methods developed in Section \ref{sec:XYmodel}, we are able to investigate the local convertibility property of {\em all} quantum ground states in the ordered phase. In Fig.~\ref{convertibilityversusell} we report the behavior of the R\'enyi entropies $S_\alpha$ as functions of the different ground states for a bipartition of the system in which subsystem $A$ is made of $\ell$ contiguous spins, while in Fig.~\ref{convertibilityversusr} we report it for subsystem $A$ made of two spins with various inter-spin distances.

We observe that the MSBGSs are the ground states characterized by the smallest value of all R\'enyi entropies, independently of the size $\ell$ of the
subsystem and of the inter-spin distance $r$. Therefore, all elements in the ground space are always locally convertible to a MSBGS, while the opposite is impossible. This first quantitative criterion for classicality is thus satisfied only by MSBGSs.

\subsection{Many-body entanglement distribution}

We now compare symmetry-breaking and symmetry-preserving ground states with respect to entanglement distribution. The monogamy inequality quantifies in a simple and direct way the limits that are imposed on how bipartite entanglement may be shared among many parties~\cite{Coffman2000,Osborne2006}. For a given many-body system of $N$ $1/2$-spins it reads:
\begin{equation}
\tau (i|N-1) \geq \sum_{j \neq i} \tau(i|j) \; \; \; \; , \; \; \; \forall \; i \; .
\label{monogamy}
\end{equation}
In the above expression, $\tau = C^{2}$ is known as the tangle, where $C$ is the concurrence~\cite{Hill1997,Wootters1998}; the sum in the r.h.s. runs over all $N-1$ spins excluding spin $i$. The l.h.s. quantifies the bipartite entanglement between one particular, arbitrarily chosen, spin in the collection (reference spin $i$) and all the remaining $N-1$ spins. The r.h.s. is the sum of all the pairwise entanglements between the reference spin and each of the remaining $N-1$ spins. The inequality implies that entanglement cannot be freely distributed among multiple quantum parties $N \geq 3$, a constraint of quantum origin with no classical counterpart.

The residual tangle $\tilde{\tau}$ is the positive semi-definite difference between the l.h.s and the r.h.s in Eq.~(\ref{monogamy}). It measures the amount of entanglement not quantifiable as elementary bipartite spin-spin entanglement. Its minimum value compatible with monogamy provides yet another quantitative criterion for classicality.

Specializing, for simplicity but without loss of generality, to translationally-invariant $XY$ spin systems in magnetically ordered phases, since the expectation value of $\sigma_i^y$ vanishes on every element of the ground space, the expressions of the tangle $\tau$ and the residual tangle $\tilde{\tau}$ for any arbitrarily chosen spin in the chain read, respectively,
\begin{eqnarray}
\tau & = & 1- m_z^2-(u^*v+v^*u)^2 m_x^2 \; , \label{tangle} \\
\tilde{\tau} & = & \tau - 2 \sum_{r=1}^{\infty} C_{r}^2(u,v) \geq 0 \label{residualtangle} \; ,
\end{eqnarray}
where $m_z\!=\!\langle e| \sigma_i^z | e \rangle\!=\!\langle o| \sigma_i^z | o \rangle$ is the on-site magnetization along $z$, the order parameter
$m_x\!=\!\langle e| \sigma_i^x | o \rangle\!=\! \sqrt{\lim\limits_{r \to \infty} \langle e| \sigma_i^x \sigma_{i+r}^x| e \rangle}$, and
$C_{r}(u,v)$ stands for the concurrence between two spins at a distance $r$ when the system is in any one of the possible ground states $|g(u,v)\rangle$,
Eq.~(\ref{eq:groundstates}).

As already mentioned, by comparing the symmetric ground states with the MSBGSs, the spin-spin concurrence is larger in the MSBGSs~\cite{Osterloh2006} if $h < h_f < h_c$, where $h_f = \sqrt{1 - \gamma^2}$ is the factorizing field, while for $h_f < h < h_c$ they are equal. In fact, we have verified that these two results are much more general. We have compared all ground states (symmetric, partially symmetry breaking, and MSBGSs) and we have found that for $h < h_f < h_c$ the spin-spin concurrences are maximum in the MSBGSs for all values of the inter-spin distance $r$, while for $h_f < h < h_c$ and for all values of $r$ they are independent of the superposition amplitudes $u$ and $v$ and thus acquire the same value irrespective of the chosen ground state. 

Finally, it is immediate to see that the third term in the r.h.s. of Eq.~(\ref{tangle}) is maximized by the two MSBGSs. Collecting all these results, it follows that the many-body, macroscopic multipartite entanglement, as quantified by the residual tangle, is minimized by the two MSBGSs and therefore also this second quantitative criterion for classicality is satisfied only by the MSBGSs among all possible quantum ground states.

\section{Conclusions and outlook}\label{sec:conclusions}

In the present work we have investigated the classical nature of globally ordered phases associated to nonvanishing local order parameters and spontaneous symmetry breaking. We have put on quantitative grounds the long-standing conjecture that the maximally symmetry-breaking ground states (MSBGSs) are macroscopically the most classical ones among all possible ground states. We have proved the conjecture by introducing and verifying two independent quantitative criteria of macroscopic classicality. The first criterion states that all global ground states in the thermodynamic limit are locally convertible to MSBGSs, i.e. by applying only local operations and classical communication (LOCC transformations), while the opposite is impossible. The second criterion states that the MSBGSs are the ones that satisfy at its minimum the monogamy inequality for globally shared bipartite entanglement and thus minimize the macroscopic multipartite entanglement as quantified by the residual tangle. We have thus verified that, according to these two criteria, the MSBGSs are indeed the most classical ones among all possible quantum ground states.

These findings lend a strong quantitative support to the intuitive idea that the physical mechanism which selects the MSBGSs among all possible ground states at the macroscopic level is due to the unavoidable presence of environmental perturbations, such as local fields, which in real-world experiments necessarily drive the system onto the most classical among the possible ground states via decoherence. This reasoning is strengthened by the fact that local perturbations may be described by LOCC transformations and for each set of parameters consistent with a globally ordered phase all quantum ground states are always locally convertible into the MSBGSs.

The above conclusions are further strengthened by the results appeared recently in Ref.~\cite{Hamma2016}, where it is shown that the MSBGSs of quantum many-body Hamiltonians have vanishing total correlations between macroscopically separated regions, as measured by the quantum mutual information.

{\em Acknowledgments} - The authors acknowledge financial support from the Italian
Ministry of Scientific and Technological Research under the PRIN 2010/2011 Research Fund, and from the EU FP7 STREP Project EQuaM, Grant Agreement No. 323714. SMG acknowledges financial support from the Austrian Science Foundation, Grant FWF-P23627-N16.

\addcontentsline{toc}{chapter}{Bibliography}

\bibliographystyle{apsrevfixed}
\bibliography{database}

\end{document}